\documentclass[reprint,superscriptaddress,showpacs]{revtex4}
\usepackage{epsf,amssymb,subfig,amsmath,color,times,natbib}
\usepackage[pdftex]{graphicx} 
\usepackage{amsmath}
\usepackage{mathtools}

\begin{document}
\title{Instability in electromagnetically driven flows\\
  Part II}

\author{Paola Rodriguez Imazio \& Christophe Gissinger} \affiliation{Laboratoire de Physique
  Statistique, Ecole Normale Superieure, CNRS, 24 rue Lhomond, 75005
  Paris, France}

\newcommand{\bfnabla}{\boldsymbol{\nabla}} 
\newcommand{\bB}{\boldsymbol{B}}                   
\newcommand{\bj}{\boldsymbol{j}}                     
\newcommand{\bu}{\boldsymbol{u}}                  
\newcommand{\btimes}{\boldsymbol{\times}}   
\newcommand{\remark}[1]{{\color{red}\bf #1}} 

\begin{abstract}
In a previous paper, we have reported numerical simulations of the
MHD  flow driven by a travelling magnetic field (TMF) in an
annular channel, at low Reynolds number. It was shown that the stalling of such induction pump is strongly related
to magnetic flux expulsion. In the present article, we show that for larger hydrodynamic
Reynolds number, and with more realistic boundary conditions, this
instability takes the form of a large axisymmetric vortex flow in the
($r,z$)-plane, in which the fluid is locally pumped in the direction
opposite to the one of the magnetic field. 
Close to the marginal stability of this vortex flow, a low-frequency pulsation  is generated. Finally, these results are compared to theoretical predictions and are discussed within the framework of experimental annular linear induction
electromagnetic pumps.
\end{abstract}
\maketitle

\section{Introduction}
\label{sec:intro}
Electromagnetic Linear Induction Pumps (EMPs) are largely used in
secondary cooling systems of fast breeder reactor, mainly because of
the absence of bearings, seals and moving parts. In these EMPs, the
conducting fluid is generally driven in a cylindrical annular channel
by means of an external travelling magnetic field . In such
induction pumps, the electrical current is induced by the variation of
the magnetic flux of the wave rather than imposed into the fluid by
electrodes, as in conduction pumps.

Nowadays, it is known that these pumps face a number of problems as
they become large enough. In particular, a strong low frequency
pressure pulsation associated to a consequent decrease of the pump
efficiency takes place at large magnetic Reynolds number $Rm$. It has
been suggested that this behavior may be related to some
magnetohydrodynamic instability. One of the first theoretical approach
to this problem was done by Galitis and Lielausis
\cite{Galaitis76}, who derived a criterion based on the magnetic
Reynolds number for the appearance of such instability. In this model,
instability arises and takes the form of an inhomogeneity in the
azimuthal direction, for sufficiently large $Rm$. This instability may
be related to experimental results obtained by \cite{Kirillov80},
\cite{Karasev89} who showed that when $Rm>Rm_c$, a low frequency
pulsation in the pressure and the flow rate is indeed observed.

More recently, significant progress have been done on the understanding
of these electromagnetically driven flows. First, it has been shown
through numerical and experimental studies \cite{Araseki00} that even
at low $Rm$, the efficiency of such pumps is affected by an
amplification of the electromagnetic forcing, which takes the form of
a strong pulsation at double supply frequency (DSF). Second, it has
been confirmed that at large $Rm$, an azimuthal non-uniformity of the
applied magnetic field or of the sodium inlet velocity can create
some vortices in the annular gap \cite{Araseki04}. In both cases
(large and low $Rm$), some solutions have been proposed to inhibit the
occurrence of these perturbations, but generally imply a strong loss of
efficiency of the pump.

In the first part of this article, we reported numerical simulations
of a very idealized configuration. We chose to model a pump infinite
in the axial direction, and at relatively small kinetic Reynolds number ($Re\sim 100$). This
allowed us to emphasize the physical mechanism by which the flow may
become unstable in such systems. In particular, we have shown that
magnetic flux expulsion and reconnection seems to control the
transition from synchronous flows to stalled regimes.

In the present part, we report a numerical study of a more realistic
configuration reproducing an electromagnetic pump. In particular, we
simulate flows at much larger fluid Reynolds number, and take into
account realistic boundary conditions in the axial direction. A new instability is reported, in which large scale vortices are generated in the flow due to MHD effects, but only when the kinetic Reynolds number 
of the flow is high enough.

We show that this instability, intrinsically axisymmetric, is related
to end-effects and can be simply understood within the framework of
classical MHD-machine theory. Although the structure and the destabilization of the flow seem very different from the stalling instability observed in the laminar regime, we will see that the mechanism reported here is similar to what has been described in the first part, except that it occurs locally in the pipe.  

We also show that complex behaviors can
arise in these electromagnetically driven flows, such as slow periodic
modulation of the flow rate.
\section{Model}
\label{sec:model}
\begin{figure*}
\centerline{\includegraphics[width=10cm,height=6cm]{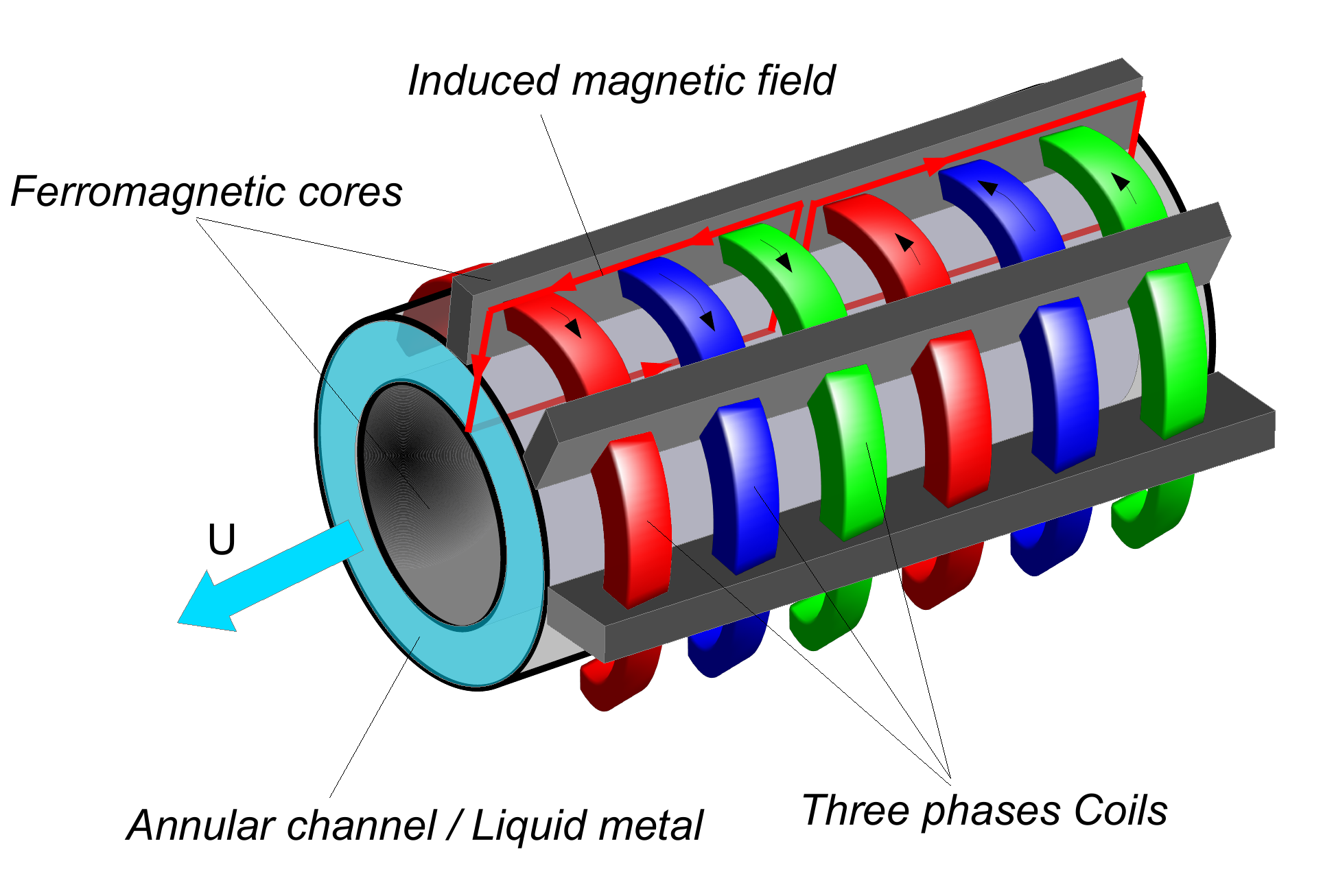}}
\caption{Schematic view of a typical linnear induction annular 
electromagnetic pump.}
\label{fig:fig1}
\end{figure*}

A schematic view of a typical electromagnetic pump is shown in
Fig.\ref{fig:fig1}. In general, the liquid metal flows along an annular
channel, between two concentric cylinders. A ferromagnetic core is
placed on the inner side of the channel, in order to reinforce the
radial component of the magnetic field created by a three-phase system
of electrical currents imposed on the outer side of the channel.

In our numerical simulations, we consider the flow of an electrically
conducting fluid between two concentric cylinders. $r_i$ is the radius
of the inner cylinder, $r_o=r_i/\beta$ is the radius of the outer
cylinder, and $H$ is the length of the annular channel between the
cylinders. In all the simulations reported here, 
periodic boundary conditions are used in the axial direction.

The governing equations are the magnetohydrodynamic (MHD) equations,
i.e. the Navier-Stokes equation coupled to the induction equation :
\begin{equation}
\rho\frac{\partial \bu}{\partial t} + \rho\left( \bu\cdot\bfnabla\right)  
\bu=-\bfnabla P+\rho\nu\bfnabla^{2}\bu +\bj\btimes\bB,
\label{eq:NS}
\end{equation}
\begin{equation}
\frac{\partial\bB}{\partial t}=\bfnabla\btimes\left(\bu\btimes\bB\right)+\frac{1}{\mu_{0}\sigma}\bfnabla^2\bB,
\label{eq:ind}
\end{equation}
where $\rho$ is the density, $\nu$ is the kinematic viscosity, $\sigma$
is the electrical conductivity, $\mu_0$ is the magnetic permeability, $\bu$
is the fluid velocity, $\bB$ is the magnetic field, and $\bj=\mu_0^{-1}
\bfnabla\btimes\bB$ is the electrical current density.  

On the cylinders, we consider infinite magnetic permeability boundary
conditions, for which the magnetic field is forced to be normal to
each boundary. In addition, an azimuthal electrical current $J_\theta$
is imposed on the outer cylinder (see \cite{part1} for more
details). This external electrical current $J$ is imposed as:
\begin{equation}
J=J_0\sin\left(kz-\omega t\right)
\end{equation}
where $J_0$ is the amplitude of the applied surface current, $\omega$ and
$k=2\pi/\lambda$ are respectively the pulsation and the wavenumber of
the magnetic field.

Our equations are made dimensionless by a  length scale
$l_0=\sqrt{r_i(r_o-r_i)}$ and a  velocity scale $u_0=c$, where
$c=\omega/k$ is the speed of the TMF. The
pressure scale is $p_0=\rho c^2$ and the scale of the magnetic field is
$B_0=\sqrt{\mu\rho}c$.

The problem is then governed by the geometrical parameters
$\Gamma=H/(r_o-r_i)$ and $\beta=r_i/r_o$, and the following
dimensionless numbers: the magnetic Reynolds number $Rm=\mu\sigma
cl_0$, the magnetic Prandtl number $Pm=\nu/\eta$, and
the Hartmann number, which controls the magnitude of the applied
current, defined as $Ha=\mu_0J_0l_0/\sqrt{\mu_0\rho\nu\eta}$.
Alternatively, one may define a kinetic Reynolds number $Re=Rm/Pm$ instead of $Pm$.

These equations are integrated with the HERACLES code
\cite{Gonzales06}, described in the first part of this
article. Typical resolutions used in the simulations reported in this
article are $(N_r,N_Z)=[256,1024]$. For the velocity field, no-slip
conditions are used at the radial boundaries. Depending on the
simulations, we can either impose an inlet velocity $U_{in}$ at $z=0$,
or an applied pressure gradient $P_{in}$ over the whole pump.

In our previous article, we studied a strongly idealized configuration
with small Reynolds numbers ($Re=100$) and axially infinite
channels. In the simulations reported in the present paper, we
explore a much more turbulent configuration, with Reynolds numbers
ranging from $Re=3000$ to $Re=10000$. In addition, we take into
account the so-called pump inlet/outlet conditions. This means that
the boundary electrical currents are applied only on one half of the
computational domain (from $z=H/4$ to $z=3H/4$), so the pump has a
finite size and discharges into a non-magnetized channel.
\section{Counter-flow instability}
\label{sec:rev}
\begin{figure*}
\centerline{\includegraphics[width=13cm,height=5cm]{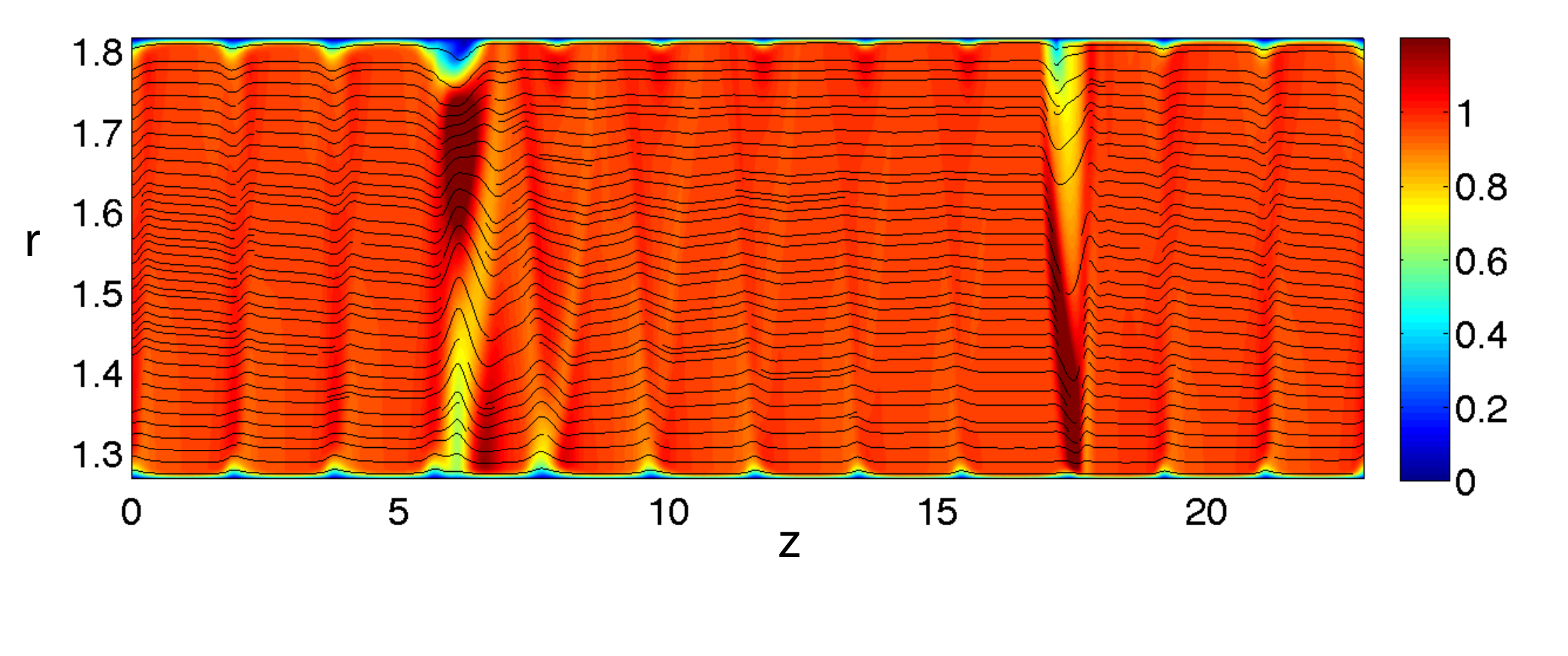}}
\vspace{-0.5cm}
\centerline{(a)}
\centerline{\includegraphics[width=13cm,height=5cm]{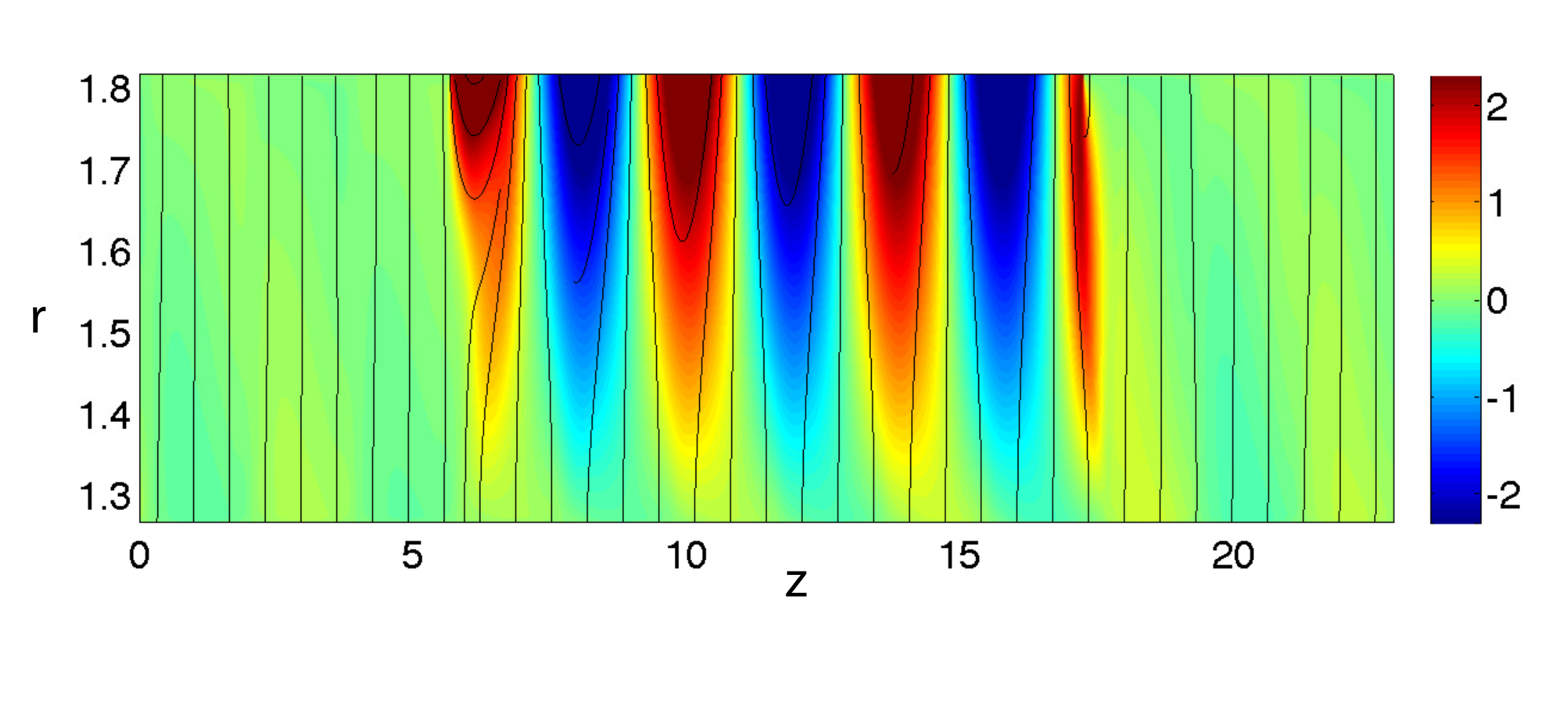}}
\vspace{-0.5cm}
\centerline{(b)}
\caption{(a) Colorplot of the velocity field and (b) of the magnetic field
for $Rm=30$, $Ha=1200$ and $Re=5000$. Streamlines and magnetic field 
lines are shown in black. Figure has been enlarged in the radial direction.}
\label{fig:fig2}
\end{figure*}

Figure~\ref{fig:fig2} shows a typical numerical simulation obtained
for $Re=5000$, $Rm=30$ and $Ha=1200$. Figure~\ref{fig:fig2}(a) shows
an instantaneous snapshot of the structure of the axial velocity field,
whereas Fig.~\ref{fig:fig2}(b) shows the instantaneous Laplace force. 
For these values of the parameters, the normalized
flow rate is close to unity in most of the domain, and the bulk of the
fluid is nearly pumped in synchronism with the magnetic field. Contrary to the axially infinite laminar pump studied in
\cite{part1}, the velocity profile (even time-averaged) is not
independent of $z$. Since the magnetic field is applied on the outer
cylinder only between $z=H/4$ and $z=3H/4$, the inlet/outlet
conditions lead to strong flow perturbations, with a stronger
effect at the inlet boundary, as shown in
Fig.~\ref{fig:fig2}.

\begin{figure*}
\centerline{\includegraphics[width=13cm,height=5cm]{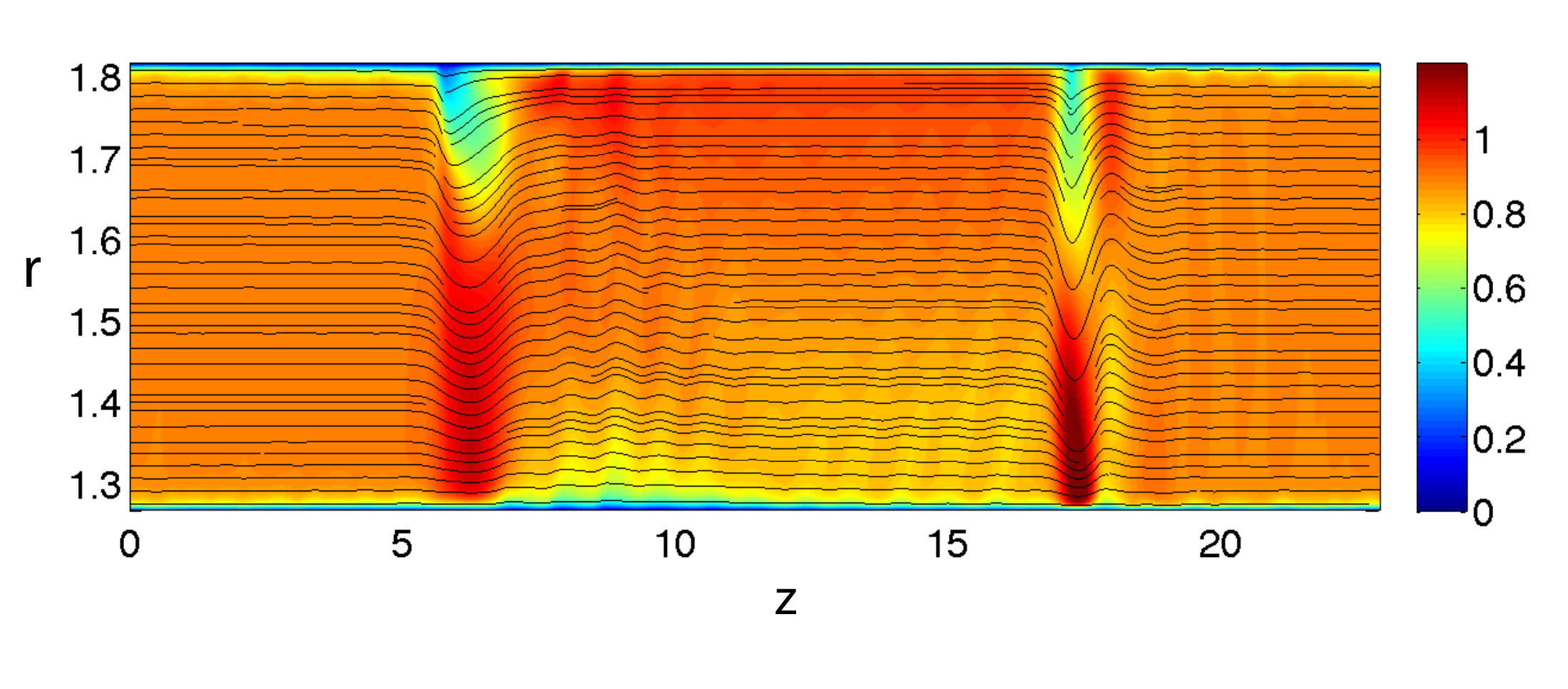}}
\vspace{-0.5cm}
\centerline{(a)}
\centerline{\includegraphics[width=8cm]{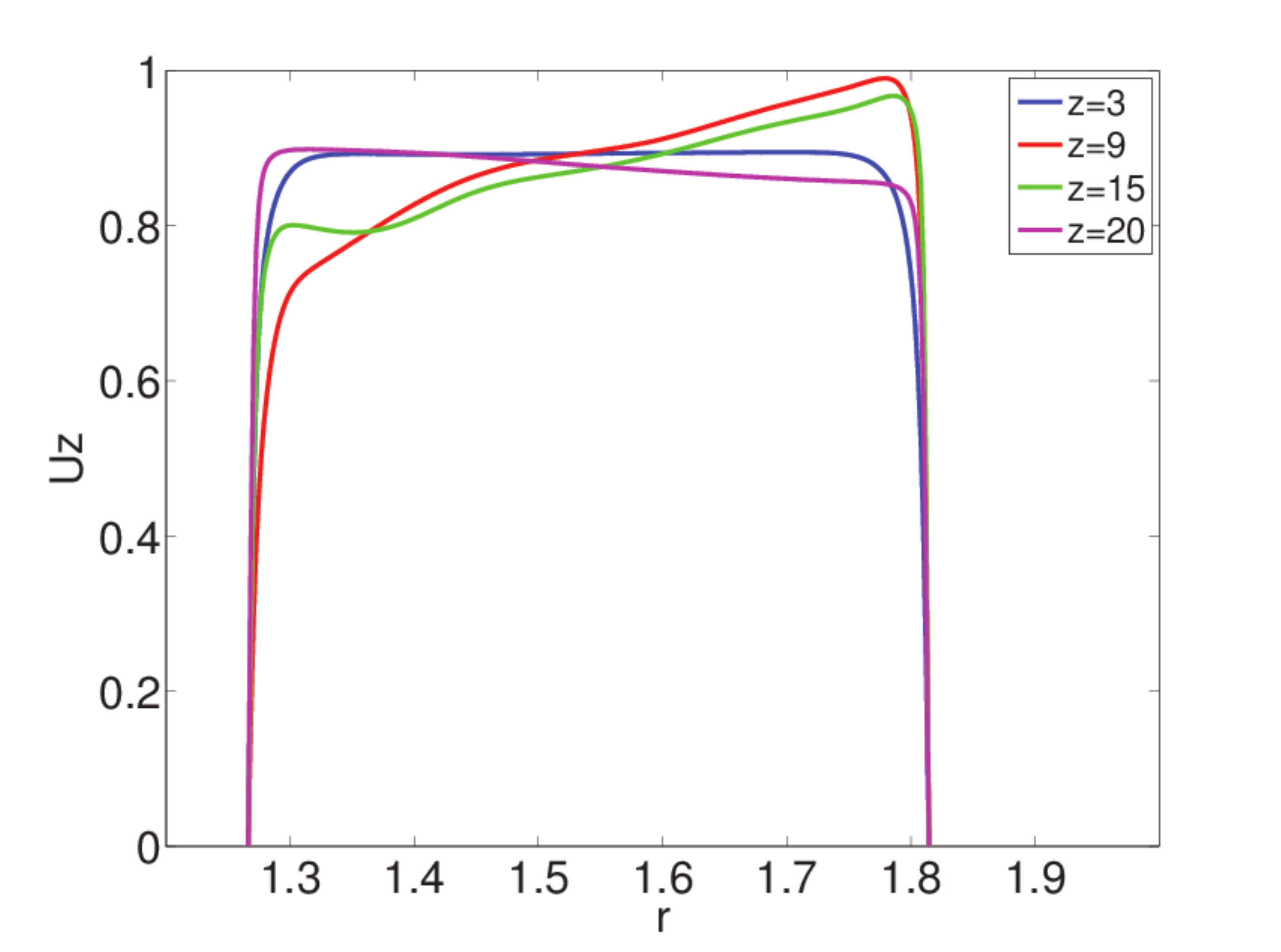}}
\vspace{-0.1cm}
\centerline{(b)} 
\vspace{-0.3cm}
\caption{(a) Colorplot of the time averaged axial velocity field for run with 
  $Rm=60$, $Ha=1200$ and $Re=5000$.
 (b) Corresponding  profiles in the radial direction for  
  different values of $z$.}
\label{fig:fig3}
\end{figure*}

In this region, the radial magnetic field is also more complex and
tends to decay with the distance from the external cylinder. 
Another strong difference with simulations performed at smaller $Re$ is the
behavior of the boundary layers. When the flow enters the active
region, there is a widening of the magnetic boundary layer close to
the inner cylinder, whereas a narrowing is observed close to the
external boundary.
Fig.~\ref{fig:fig3} shows the time-averaged velocity field for a
larger magnetic Reynolds number ($Rm=60$) and illustrates the spatial
organization of the flow. Before entering the active region ($z<H/4$),
the flow exhibits a laminar structure relatively symmetrical and
identical to the solutions obtained at smaller $Re$ and reported in
the first part of the paper. Despite the absence of applied magnetic
field in this region, note that the velocity profile is different from
a classical annular Poiseuille flow. The presence of a relatively flat
profile in the bulk flow and a stronger shear near boundaries is more
typical of magnetized regimes.

As the velocity field is probed at larger $z$, inside the active
region ($z=9$ and $z=15$), the maximum value of $U_z$ shifts towards
the outer cylinder, leading to stronger velocity gradient close to
this boundary. On the contrary, the inner boundary layer widens. This effect is not observed at smaller
$Rm$, where the velocity (except in the boundary layers) is strongly
homogeneous in the radial direction. Finally, outside the active
region, the system comes back very rapidly to the symmetrical
profile with magnetic boundary layers on each sides of the annular
gap.

In addition, there is a strong perturbation as the fluid enters or
leaves the active region of the electromagnetic pump. These
perturbations at the inlet/outlet boundaries induce a large adverse
pressure gradient inside the channel and can yield local velocities
stronger than the wave speed.\\

It is important to note that despite strong fluctuations, the flow for
$Rm=30$ and $Rm=60$ stays relatively homogeneous in most of the
computational domain, and the total flow rate developed by the
channel is very close to its maximum value. 

\begin{figure*}
\centerline{\includegraphics[width=13cm,height=5cm]{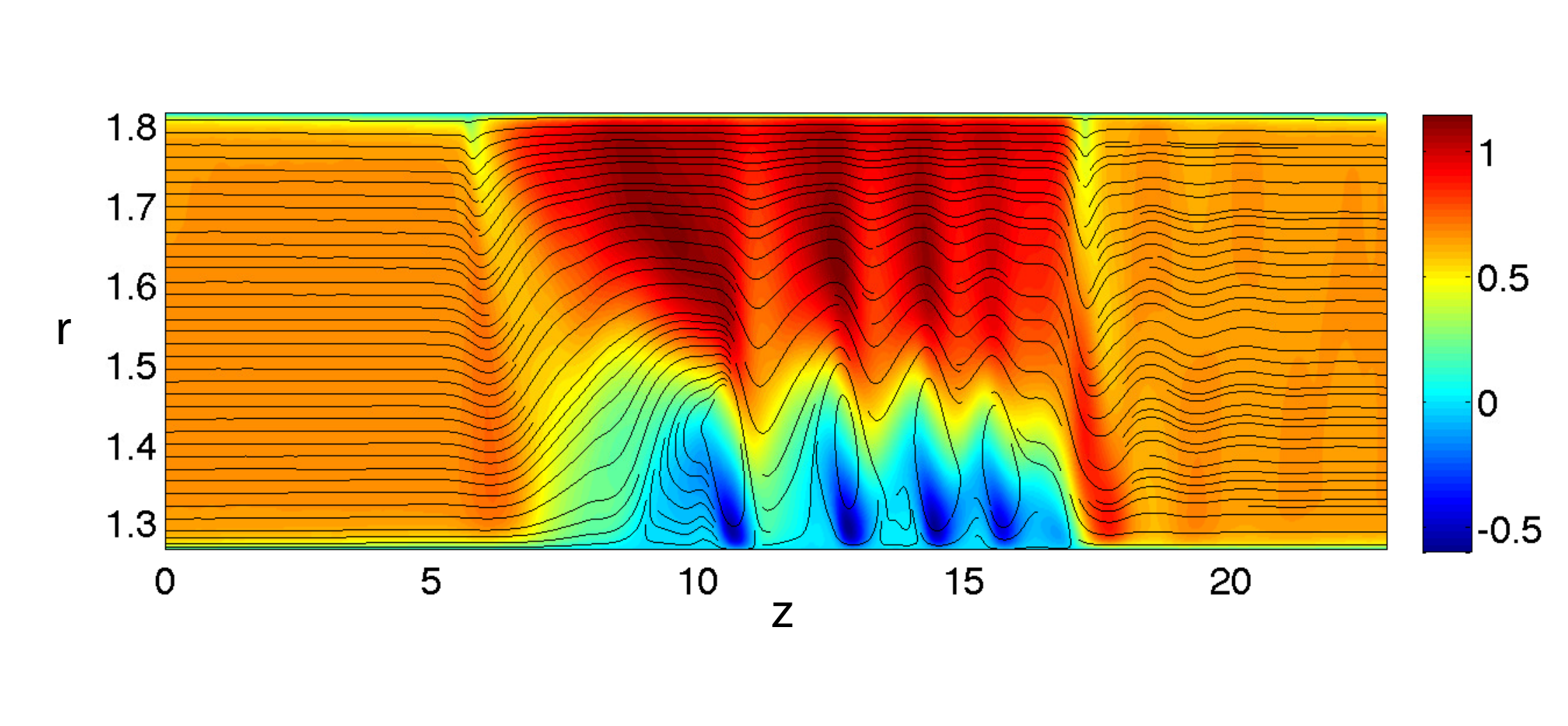}}
\vspace{-0.5cm}
\centerline{(a)}
\includegraphics[width=7cm,height=6cm]{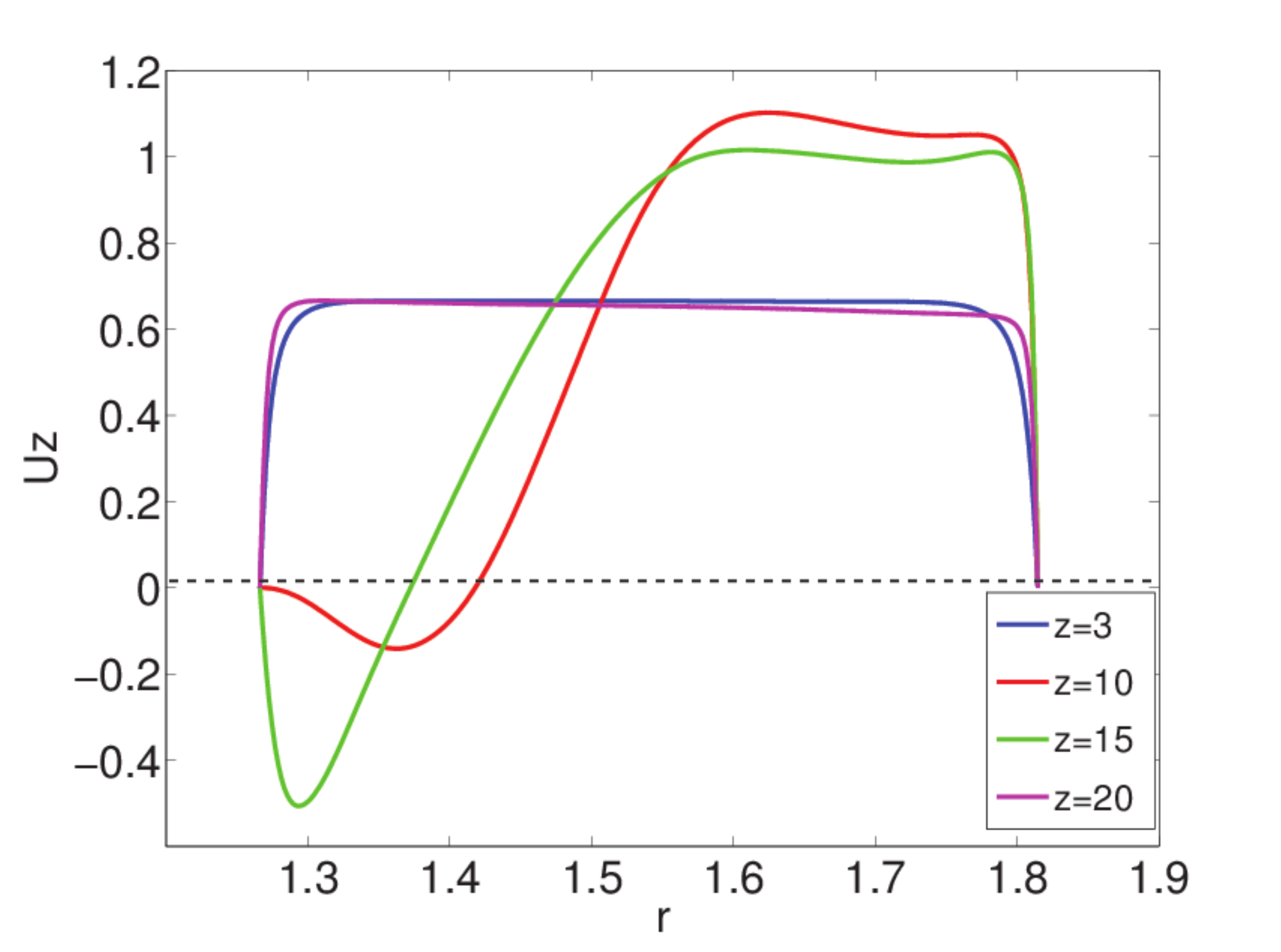}\vspace{-0.5cm} \includegraphics[width=7cm,height=6.2cm]{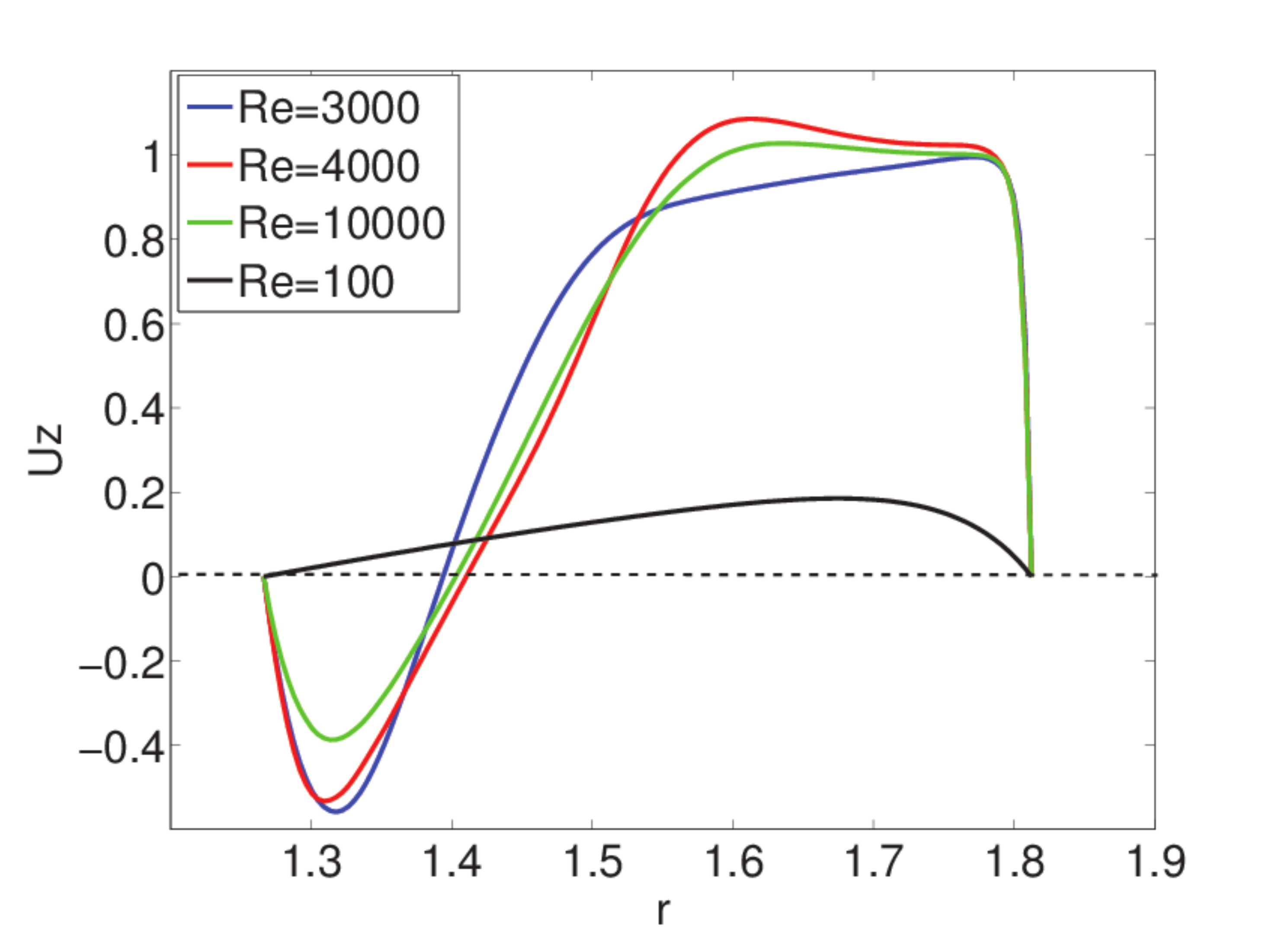} 
\centerline{(b)} 
\vspace{-0.3cm}
\caption{(a) Colorplot of the time averaged velocity field for run
  with $Rm=120$, $Ha=1200$ and $Re=5000$.  (b) Corresponding profiles
  in the radial direction for different values of $z$. (c) Velocity
  profiles in the radial direction for runs with $Re=3000, 4000$ and
  $10000$, for fixed $Rm=120$, $Ha=1000$. Note the difference with
  $Re=100$ (obtained for $Rm=120$, $Ha=100$). }
\label{fig:fig4}
\end{figure*}

According to the solid body model for induction pumps \cite{Galaitis76},  an
instability of the flow similar to the stalling of an asynchronous
motor should appear at sufficiently large $Rm$. Our numerical simulations at low $Re$ exhibit a behavior very close to the predictions. The magnetic Reynolds number is therefore
further increased, from $Rm=60$ to $Rm=120$, keeping fixed $Ha=1200$
and $Re=5000$. The turbulent flow studied here develops a behavior
drastically different from the stalled regime observed in the laminar
case. Fig.~\ref{fig:fig4}(a) shows a colorplot of the averaged velocity
field in such regime. Compared to the regime obtained at lower $Rm$,
the flow rate developed by the channel has been strongly reduced with
a total normalized flow rate around $Q\sim 0.5$. This decrease in the efficiency of the pump takes the form of a strong inhomogeneous flow in both radial and axial
direction. Indeed, one can see that the fluid located close to the
outer cylinder, where the surface current is applied, moves nearly in
synchronism with the wave, similarly to the regime obtained at lower
$Rm$. On the contrary, the inner region exhibits a strongly
fluctuating {\it negative} velocity, flowing against the TMF. These negative velocities can be observed in the
time averaged velocity profiles shown in Fig.~\ref{fig:fig4}(b). Whereas the profiles
outside the pump remain Hartmann-like, inside the pump the profiles
drastically change its shape, achieving negative velocities of the
same order than the positive ones, for $r < 1.5$. For some
values of $r$, it is possible to observe positive velocities larger
than the driving wave speed. 

Fig.~\ref{fig:fig4}(c) shows the time averaged axial velocity profiles in the center
of the $z$ domain for runs at different Reynolds numbers, from $Re=100$ up to $Re=10000$, at fixed $Ha=1000$. For these parameters, the inhomogeneous regime shown in 
Fig.~\ref{fig:fig4}(a) appears for the three high $Re$ number simulations as well, 
giving similar profiles than for $Re=5000$, although very different from the laminar case
($Re=100$).

\begin{figure*}
\centerline{\includegraphics[height=8cm]{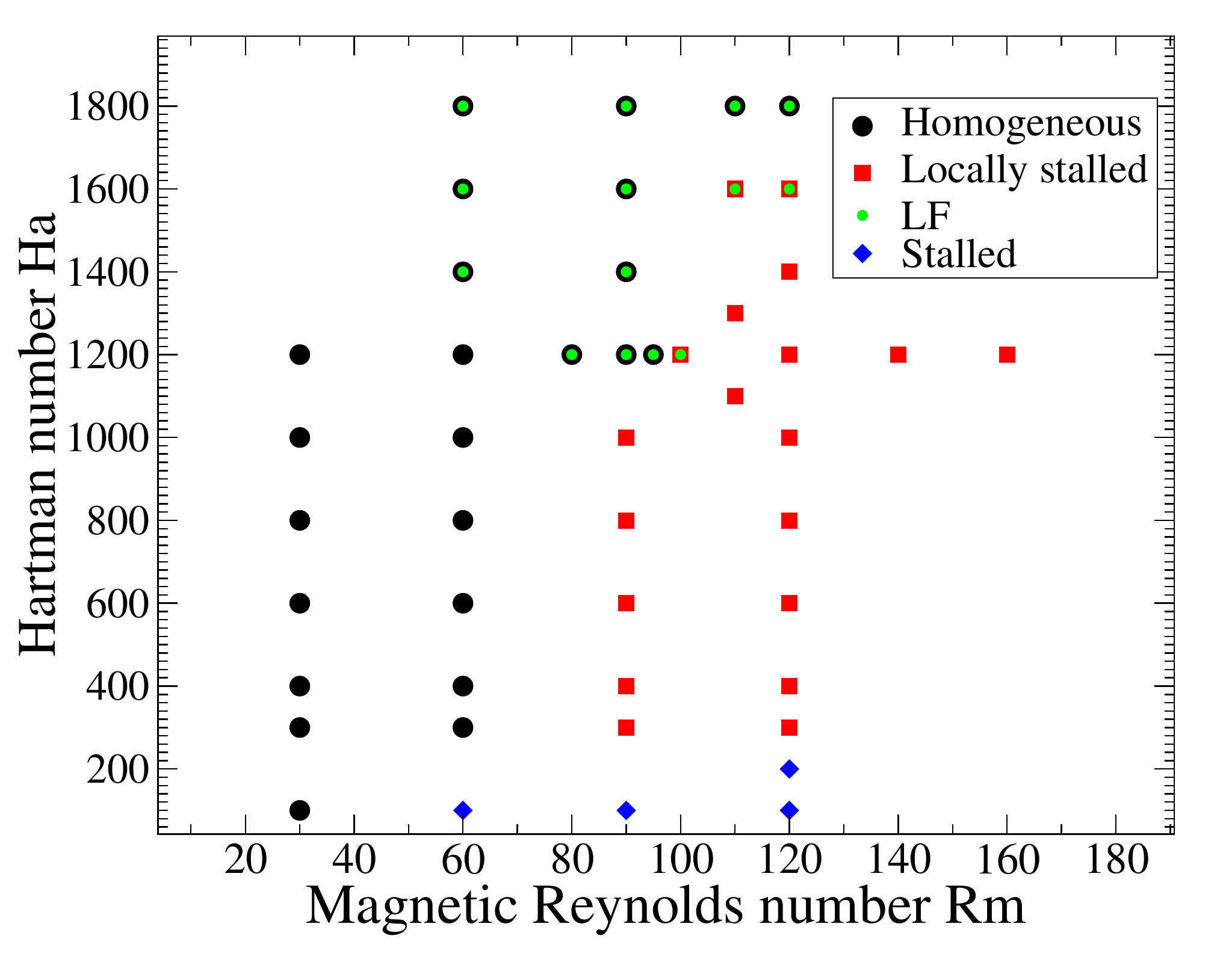} }
\caption{Parameter space $(Ha,Rm)$ explored at a given Reynolds number
  $Re=5000$. Black circles correspond to the stable solution depicted
  in Fig.~\ref{fig:fig2}, where the flows are homogeneous in $r$. Blue points 
  also correspond to flows homogeneous in $r$, but associated to very small 
  flowrates (classical stalling instability). 
  Red squares  correspond to simulations with inhomogeneous 
  flows as the ones shown in Fig.~\ref{fig:fig4}.
   Low frequency pulsation (see text), when
   present, is indicated by green circles.}
\label{fig:fig5}
\end{figure*}

A careful study of each simulation computed for different $Ha$ and
$Rm$ numbers, at a fixed value of the fluid Reynolds number $Re=5000$
gives the $Ha-Rm$ parameter space shown in
Fig.~\ref{fig:fig5}.  Black circles indicate simulations similar to
the one described in Fig.~\ref{fig:fig2} and Fig.~\ref{fig:fig3}: in
most of the domain, the fluid is relatively homogeneous, 
independent of $r$ (except close to the boundaries) 
and in synchronism with the wave.

Blue squares represent simulations where a very small flowrate is
observed. In these runs, the flow is still homogeneous in $r$, but the
velocities remain very small in comparison with the speed of the wave.
This situation is reminiscent to the one observed in the first part of
the article, associated with a global stalling of the pump.

Red squares correspond to the new state described in
fig.~\ref{fig:fig4} in which the flow becomes unstable and
inhomogeneous in $r$. 

This parameter space clearly differs from the one obtained in the case
of laminar simulations seen in part 1, for which only one transition is
observed, from homogeneous synchronism with the wave (black circles)
to homogeneous stalled flow (blue squares). The difference therefore
lies in the presence of inhomogeneous flows, characterized by the
presence of both synchronous and stalled flows, in different parts of the channel.

It also should be noted that the parameter space does not seem to
show a strong dependence on the Reynolds number, at least for $Re$
comprised between $3000$ and $10000$, as can be observed from
comparison between Figs. ~\ref{fig:fig4}(b) and (c). Nonetheless, our
simulations indicate that the upper boundary of the instability pocket
decreases as we decrease $Re$ (from $Ha\sim 1000$ to $Ha\sim 800$,
when going from $Re=5000$ to $Re=3000$), whereas the lower
boundary remains among the same values of $(Ha-Rm)$. This narrowing
of the instability pocket as we decrease $Re$ is consistent with the
theoretical predictions of the solid body model presented in the first
part of the paper. At small values of $Re$, the parameter space of the
laminar problem is divided by a single marginal
stability line following $M_b^c\propto \sqrt{Re_s}$, where $Re_s=(c-u)l_0/\nu$ is the kinetic Reynolds number based on the slip. Here, $M_b^c=(c-u)\sqrt{\rho\mu_0}/B_0$ is  the critical Alfvenic Mach number comparing the velocity of the fluid to the Alfven wave celerity at onset. It measure the ability of the fluid to expel magnetic flux from the bulk. The simulations reported in this second paper show that as $Re$ is increased, a new regime appears, opening the marginal instability line into a
pocket for which both synchronous and stalled flows coexists in the
bulk flow.  In most of the parameter space, both boundaries of this pocket
present a strong hysteresis.  In our simulations, the limit between
the three regimes shown in Fig.~\ref{fig:fig5} was always obtained
entering the instability pocket from the homogeneous solution (from
black or blue points towards red points).  In this fashion, the upper
limit of the pocket is expected to follow the turbulent scaling $M_b^c\propto 1$ (see part I) if flux expulsion is still the mechanism involved in the generation of this new instability. 

\section{Composite model}

At this point, it is interesting to compare the present high Reynolds
number results with theoretical predictions and laminar simulations
discussed in the first part of this article. First, the structure of
the parameter space shown in Fig.~\ref{fig:fig5} is very different,
with the presence of two boundaries for the generation of the
instability associated to an inhomogeneous flow (and the related
decreased flow rate). In addition, the presence of a negative flow
rate, with fluid moving {\it backward} with respect to the driving
magnetic field, seems in strong contradiction with the simple
usual description of MHD induction machines.\\

In fact, this large scale destabilization of the flow can be regarded
as an effect of a {\it local stalling} of the flow, in which different
regions of fluid (for different values of $r$) can be considered as
several elementary electromagnetic pumps. Since the total flow rate
must be equal to the sum of the flow rate in each elementary region,
and if we suppose that the applied magnetic field is almost
independent of $r$, the individual pumps are connected in parallel
hydraulically and in series electrically. In this case, each
different region is associated to identical control parameters $Rm$
and $Ha$.

The colorplot shown in Fig.~\ref{fig:fig4} suggests that only the
region close to the inner cylinder stalls, whereas the fluid located
near the outer cylinder stays in synchronism. We therefore consider
the simplest composite model containing two hydraulically parallel and
electrically independent sub-channels. Suppose that channel $1$
corresponds to the quarter of the domain close to the coils (say pump
$1$), while channel $2$ corresponds to the last quarter close to the
inner cylinder (pump $2$). Since it corresponds to a
transition zone, the central part of the fluid is not described here.

\begin{figure}
\centerline{
\includegraphics[width=7.5cm,height= 5 cm]{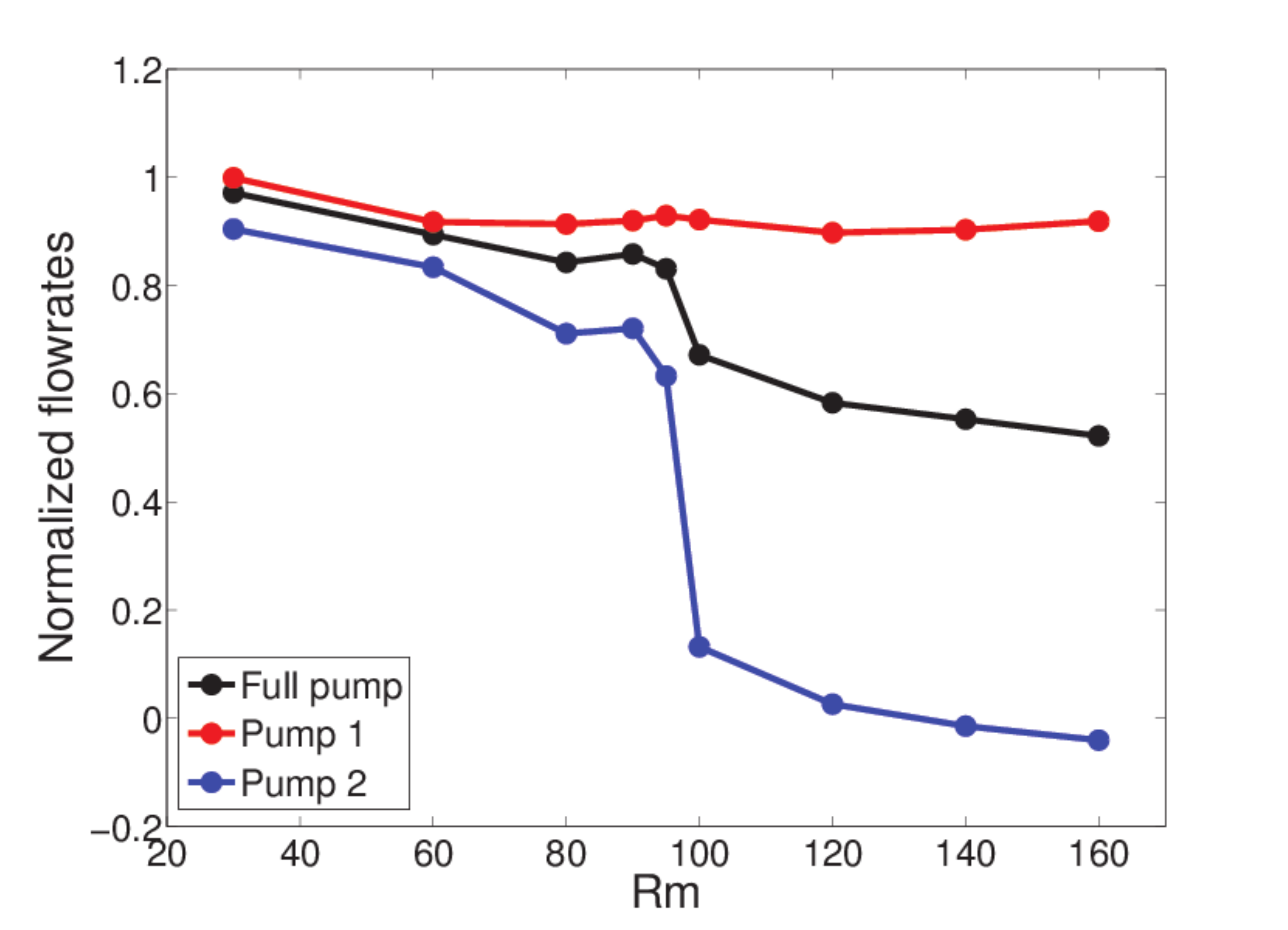}\hspace{-0.6cm} 
\includegraphics[width=7.8cm,height= 5 cm]{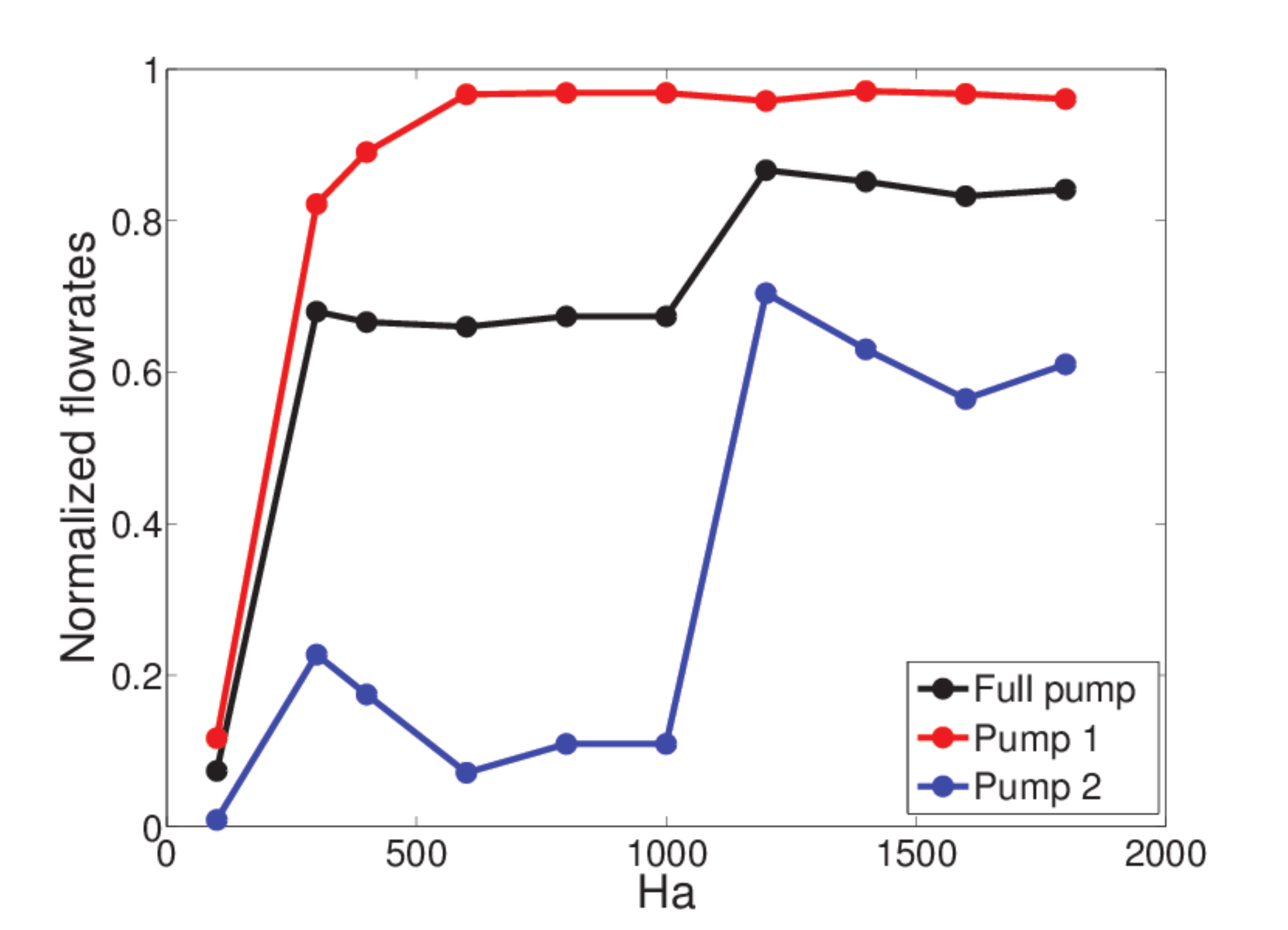}}
\vspace{-0.2cm}
\hspace{3.23cm}(a)\hspace{6.7cm}(b)
\caption{(a) Flowrate as a function of the magnetic Reynolds number
  for $Ha=1200$. (b) Flowrate as a function of the Hartmann number  
  for $Rm=90$.  In both cases, the black points represent the values
  considering the full channel, while the red and blue lines stand for
  pump $1$ and pump $2$ respectively.}
\label{fig:fig6}
\end{figure}

Figs.~\ref{fig:fig6}(a) and (b) show the evolution of the normalized
flow rate $Q_i=\frac{1}{S_i}\int_{S_i}\frac{U_z}{c}dS$ of the two elementary
pumps of section $S_i$ as a function of $Rm$ and as a function of
$Ha$, respectively, as the system enters the instability region.

\begin{figure}
\centerline{\includegraphics[height= 8cm]{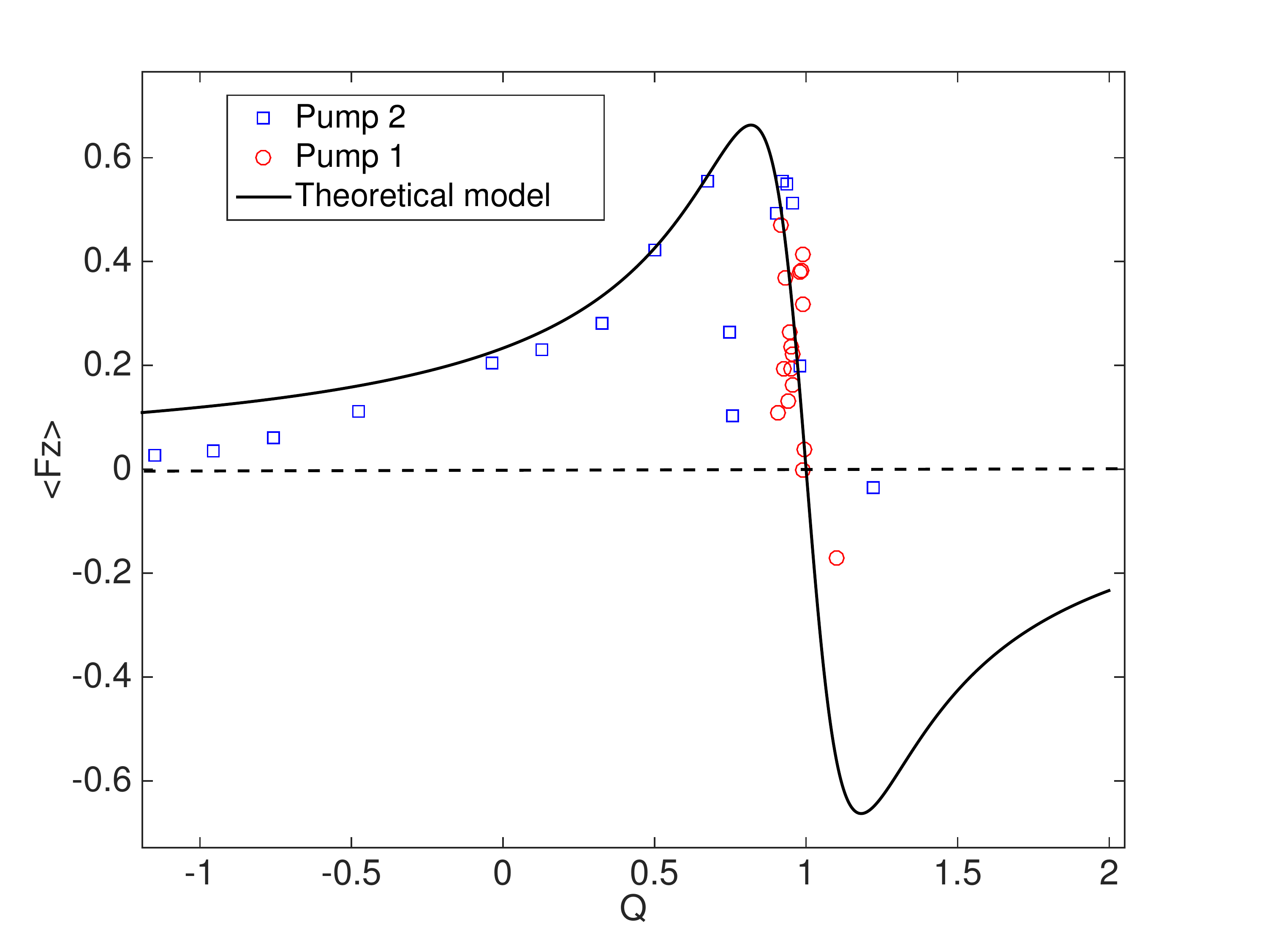}}
\vspace{-0.5cm}
\caption{PQ curve for $Rm=60$, $Ha=1000$. The black line represents the theoretical model   considering the flowrates obtained for the full channel, while the red and
  blue points stand for pump one and pump two respectively.}
\label{fig:PQcurve}
\end{figure}

At small $Rm$ and relatively large $Ha$, the total flow rate (black
curve) is nearly independent of $r$, yielding identical $Q_i$ in both
sub-channels. The total flow rate then abruptly decreases for
$Rm>Rm_c$, with $Rm_c \sim 90$ (see Fig.~\ref{fig:fig6}(a)).  The
evolution of the two elementary channels shows clearly that this
decrease is essentially due to a stalling of the pump $2$ with a flow
rate dropping to $Q_2 \sim 0.1$, whereas the flow rate of the fluid
close to the coils keeps values surprisingly close to synchronism
($Q_1>0.8$). The evolution of $Q_2$ alone is very similar to
the results obtained in low-$Re$ simulations for the whole channel. 
In particular, the local values of the magnetic Mach number $M_b$ in channel $1$ are all much smaller than one, while inhomogeneous flow are generated when local $M_b$ in channel $2$ reaches a critical value $M_b^c\sim 0.1$. This last point highlights the fact that flux expulsion is still the physical mechanism generating the inhomogeneous flow observed here, except that it only occurs locally.

If $Ha$ is decreased at fixed $Rm$ (Fig.~\ref{fig:fig6}(b)), two
transitions are observed. For $Ha<1200$ a first bifurcation occurs,
corresponding to the one described above, i.e. the stalling of the
inner region of the channel only. As $Ha$ is decreased further, a
secondary bifurcation is observed, in which the flow rate in the outer
region, close to the coils, also drops to zero. This second transition
is closer to the behavior observed for asynchronous motors, and
corresponds to a {\it global stalling} of the flow.

It is instructive to compute the typical
Pressure-Flow rate characteristic curve of the system. To do this, a
set of simulations for parameters outside the instability pocket
($Ha=1000$, $Rm=60$) was performed, imposing an increasing flowrate to
the whole system, from $Q=-1.5$ to $Q=1.5$. For each run, time and space
averaged velocity, pressure drop and Laplace force were measured
inside the whole channel, but also in each region corresponding to
the two individual pumps.

Fig.~\ref{fig:PQcurve} shows the Laplace force as a function of the
flowrate in each sub-channel, computed only for $7<z<15$, in order to
exclude the fluctuations occurring at the inlet/outlet boundaries. The
most striking result is the fact that although the $PQ$ curve for the
total channel in this turbulent regime is more complex than the
theoretical one, each elementary channel seems to lie on a single
common curve. Once again, it must be kept in mind that there is a
variation of the Laplace force with $r$ as we move along the
sub-channels, so the pumps are not fully in series electrically.  As a
consequence, it appears from Fig.~\ref{fig:PQcurve} that the fluid in region
$1$(near the coils) always moves in synchronism, while pump $2$ can
either be in synchronism (upper stable region of the parameter space)
or exhibit smaller velocity.\\

This reinforces the interpretation of two hydraulically parallel pumps,
with similar control parameters $Rm$ and $Ha$ but lying on different
regions of their characteristic. More exactly, for a given imposed
total flow rate $Q$, there is always one solution corresponding to
both sub-channels delivering identical discharge. When $Q$ is large
enough, this solution is the only one and both sub-channels flow in
synchronism with the wave (on the descending branch of the
$PQ$-curve). As the total flow rate decreases, both elementary pumps
move along their characteristic until the maximum of the curve is
reached. Below this critical value of $Q$, a second solution
corresponding to different elementary discharges is possible, the
total flow rate being the sum of the two elementary ones. In this regime,
Fig.~\ref{fig:PQcurve} shows that the functioning point of channel $1$
(red dots) is always located on the descending branch of the PQ curve,
close to synchronism, whereas only the internal part of the fluid
('pump $2$') moves on the ascending branch.  For instance, when the
imposed flow rate is $Q=0.2$, the fluid located in the inner region
exhibits a strongly negative flow rate ($Q_2=-0.6$) in order to
compensate the large positive flow rate $Q_1=0.9$ close to the coils
and satisfy incompressibility. In order to highlight this
interpretation, we added to Fig.~\ref{fig:PQcurve} the solution
expected in the framework of simple solid-body theory (eq. $2.17$ in
\cite{part1}, with $H=2.7$ and $R=5.6$).

However, there are also important differences between this simple
composite model and our numerical simulations. First, the assumption
of only two pumps in parallel is very restrictive. For instance, this
description ignore the central part of the annulus, or the boundary
layers in which the velocity vanishes. In addition, the magnetic field
being imposed only at the boundaries, the simulations involve a local
Hartmann number which depends on the radial direction, in
contradiction with the assumption of pumps connected in series
electrically. Finally, some of the numerical points in
Fig.~\ref{fig:PQcurve} significantly differ from the theoretical
curve (in particular the presence of two maxima for the Lorentz force
in channel $2$, at $Q_2\sim 0.7$ and $Q_2\sim 0.9$). This departure
finds its origin in the fact that very close to synchronism, flux
expulsion vanishes and electrical currents generated at the outer
cylinder can propagate easier through the radial gap, thus leading to
an amplification of the induced field close to the inner
cylinder. This behavior is reminiscent of the localized nature of the
forcing, which is not described by the classical solid body theory.\\

The above description is almost equivalent to the {\it composite
model} discussed in \cite{Galaitis76}. In this article, Gailitis and
Lielausis proposed that an electromagnetic pump can be modeled, in the
limit of small gap, as a combination of many elementary pumps
connected in $\phi$, therefore predicting a loss of homogeneity of the
velocity distribution along the azimuthal direction above some critical
value of the magnetic Reynolds number. In its simplest form, this
model involves two elementary pumps, similarly to the results reported
here. Even in axisymmetric configurations such as the simulations reported here, a similar
mechanism occurs, except that the loss of homogeneity is achieved in
the radial direction instead. This inhomogeneous structure, associated to a strong radial shear, is likely to produce destabilization of the flow in the azimuthal direction. Although additional 3D runs will be necessary to conclude, this destabilization would provide a new scenario, different from the mechanism proposed in \cite{Galaitis76}, for the occurence of non-axisymmetric states in such annular induction pumps.

\section{Low Frequency pulsation}
\label{sec:LF}

\begin{figure}
\centerline{
\includegraphics[width=7.8cm,height= 6 cm]{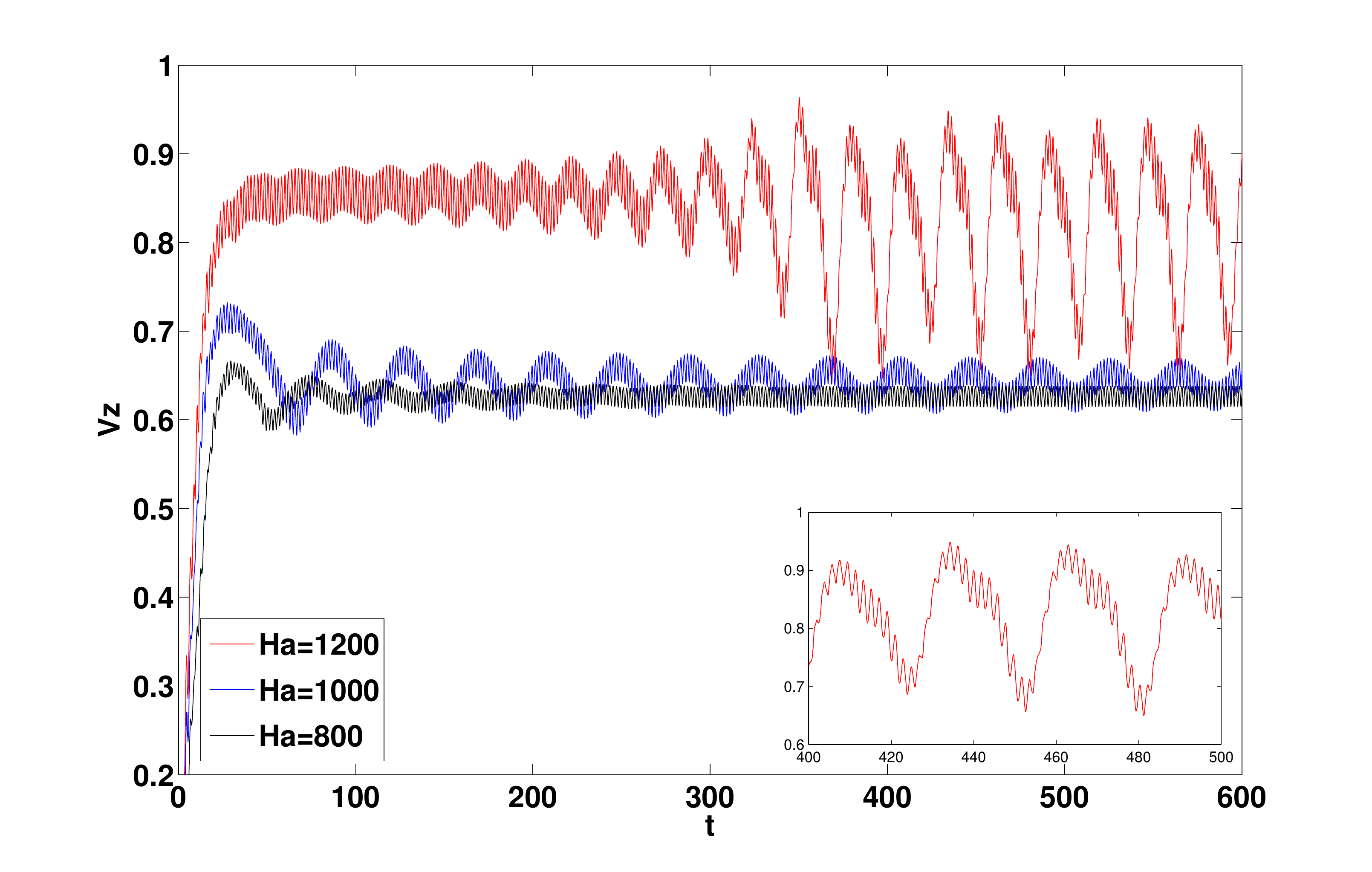} \hspace{-0.8cm}
\includegraphics[width=7.8cm,height=6 cm]{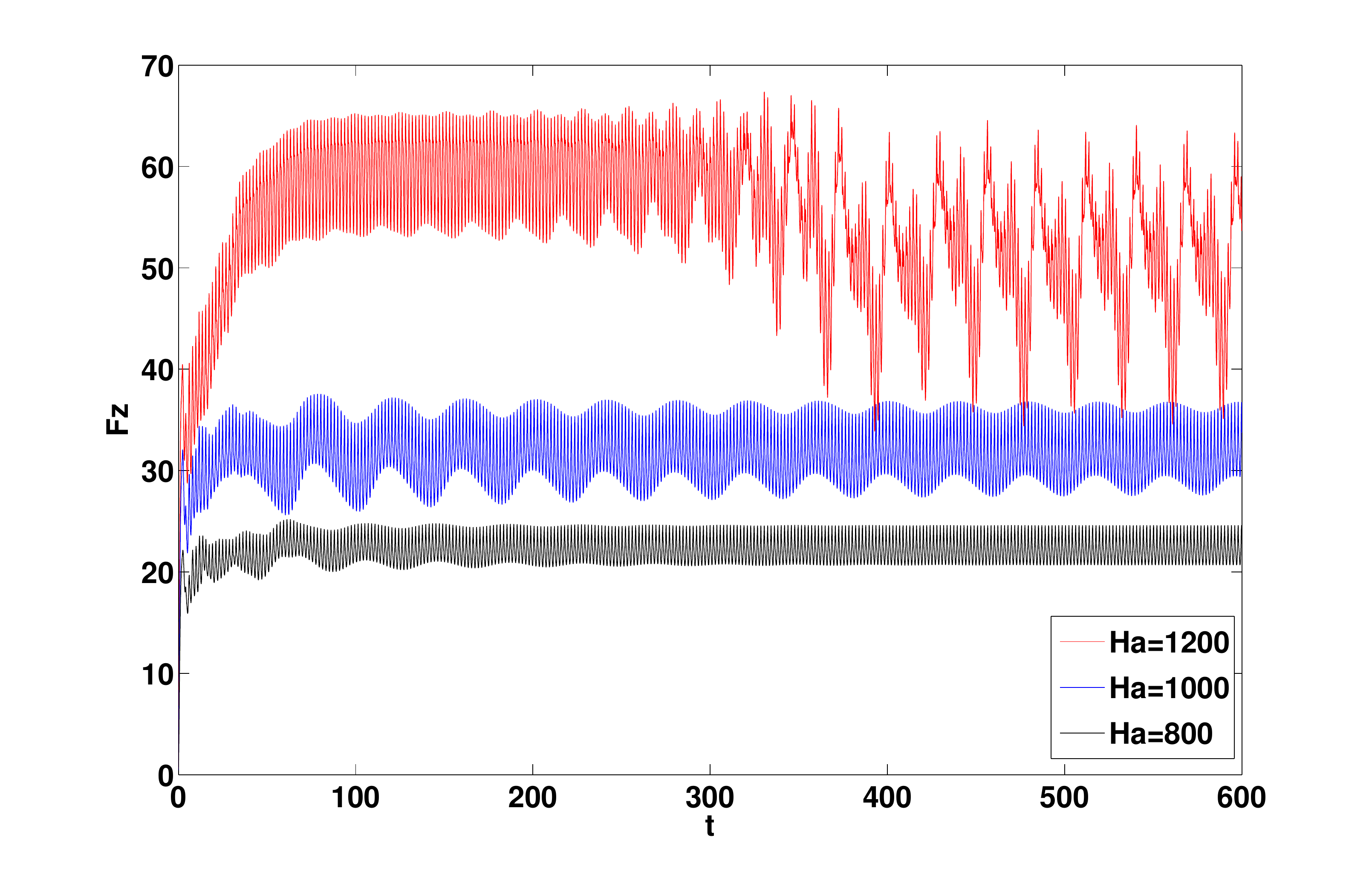}}
\vspace{-0.2cm}
\hspace{3.2cm}(a)\hspace{6.76cm}(b)
\caption{(a) Time series of the normalized axial velocity $V_z$ in the central
  part of the pump, averaged in $r$, for $Rm=90$ and various values of
  $Ha$. The inset shows an enlarged view of the pulsation for
  $Ha=1200$.  (b) The same for the Lorentz force. As $Ha$
  is increased, a low frequency pulsation develops in both fields.}
\label{fig:fig7}
\end{figure}
\begin{figure}
\centerline{\includegraphics[width=13cm,height=4.5cm]{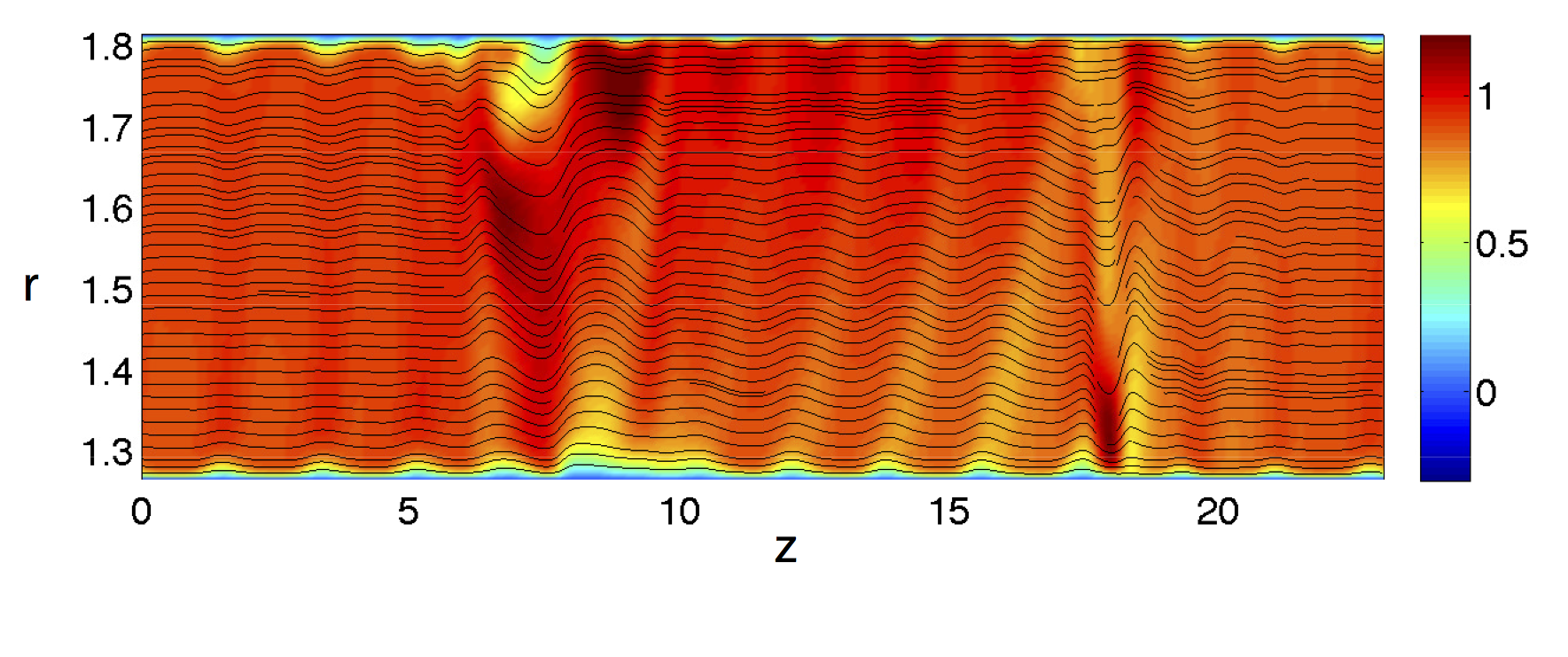}}
\vspace{-0.8cm}
\centerline{\includegraphics[width=13cm,height=4.5cm]{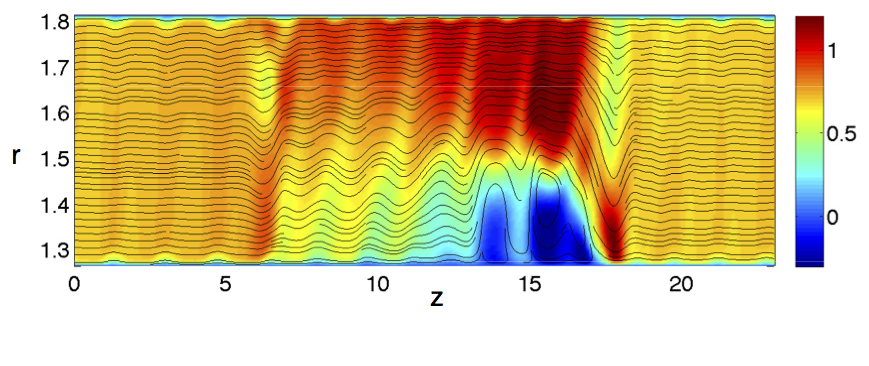}}
\vspace{-0.8cm}
\centerline{\includegraphics[width=13cm,height=4.5cm]{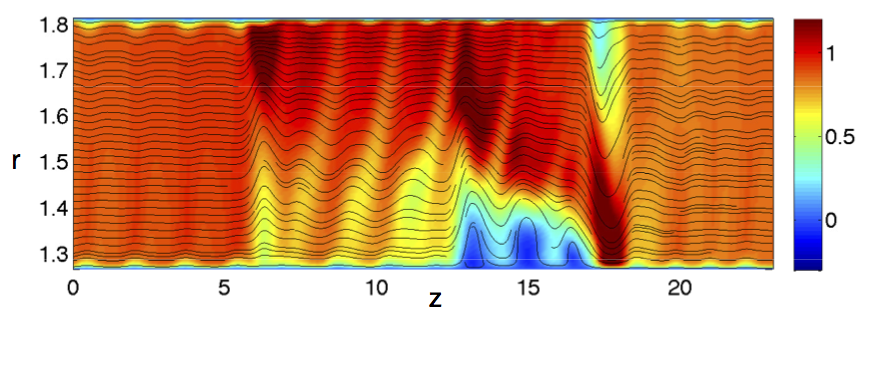}}
\vspace{-0.8cm}
\centerline{\includegraphics[width=13cm,height=4.5cm]{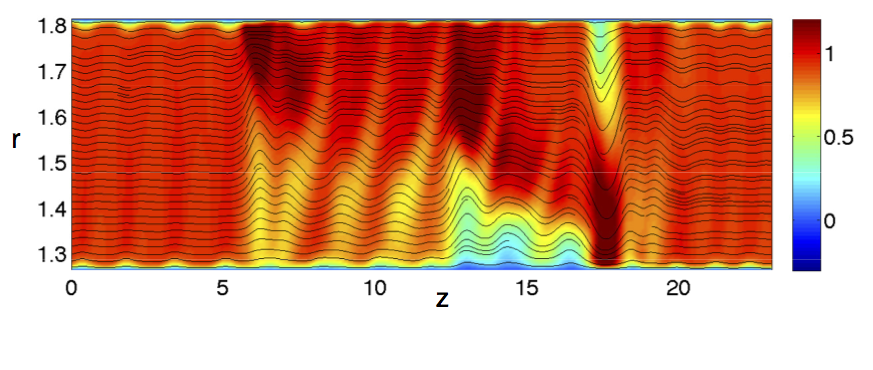}}
\vspace{-0.5cm}
\caption{Colorplot of four different snapshots of the axial velocity for run with $Rm=90$, $Ha=1200$. A vortex 
near the upper boundary appears and disappears (from top to bottom) following a cycle with $\tau \sim 30$.}
\label{fig:fig8}
\end{figure}
\begin{figure}
\centerline{
\includegraphics[width=8.3cm,height=5.5cm]{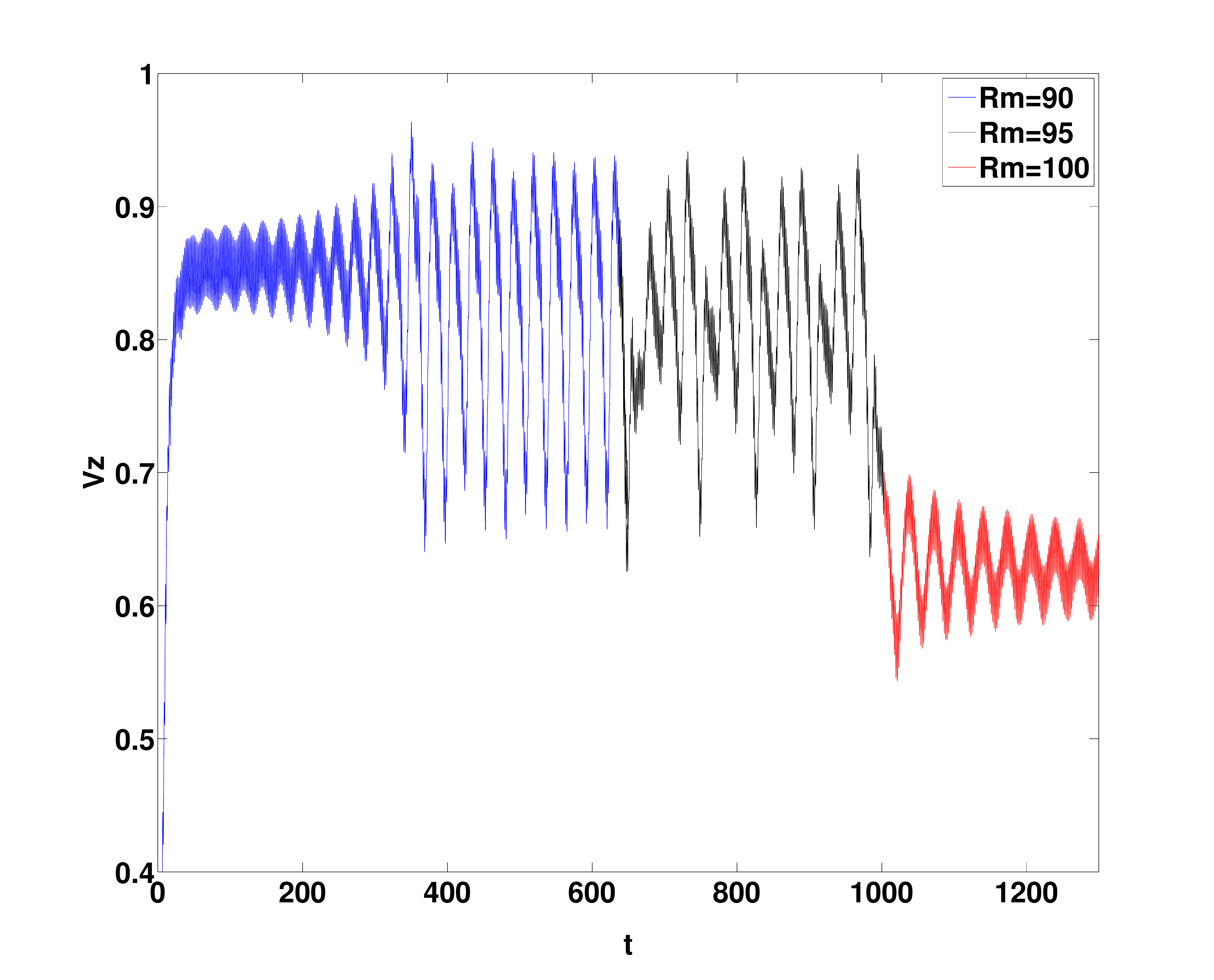}}
\caption{Time series of the axial velocity $V_z$ in the central
  part of the pump, averaged in $r$, for fixed different values of Rm.}
\label{fig:figLF}
\end{figure}

Despite strong turbulent fluctuations, the large scale vortex flow
observed in Fig.~\ref{fig:fig4} is statistically stationary when the
system is located deep inside the instability pocket (represented by
red dots in Fig.~\ref{fig:fig5}). However, a more complex time
behavior is systematically observed close to the marginal stability
line. Green symbols in Fig.~\ref{fig:fig5} indicate solutions in which
a periodic modulation of the flow rate is
obtained. Figs.~\ref{fig:fig7}(a) and (b) show the time series for
the $z$ component of velocity field and Lorentz force for runs
performed at $Rm=90$ and different values of $Ha$.
The pulsation due to the forcing, at
twice the wave frequency, is clearly visible on both fields for all
the values.  As $H$ is increased above $H=800$, a slower modulation,
associated to a typical frequency one order of magnitude larger than
the driving frequency, develops. As the applied field is increased,
the oscillation becomes strongly non-linear, as this dynamics exhibits
slowing down close to synchronism, as indicated in the inset of
Fig.~\ref{fig:fig7}(a).

As shown in Fig.~\ref{fig:fig5}, this {\it Low Frequency (LF)
  pulsation} occurs both inside and outside the instability pocket
(for instance, at $Rm=60$, $Ha=1400$, or at $Rm=120$, $Ha=1600$, as
indicated by green points in Fig.\ref{fig:fig5}), but always close to
the marginal stability line. Fig.~\ref{fig:fig8} shows snapshots of
the flow at four different times during the oscillation shown in the
inset of Fig.~\ref{fig:fig7}(a). It can be seen that this regime
corresponds to localized vortices, which periodically appear near the
pump outlet with decreasing
amplitude until they disappear.

The appearance of this regime displays a strong hysteresis, and can
exhibit more complex dynamics. Fig.~\ref{fig:figLF}(a) shows the time
evolution of the axial velocity $V_z$ as $Rm$ is increased (starting
each simulation using as initial condition the output from the
previous one) at a fixed Hartmann number $Ha=1200$. At $Rm=90$, the
pulsation appears with a single frequency roughly ten times the
frequency driving. As $Rm$ is increased beyond $Rm=95$, a second
frequency appears in the spectrum, whereas for $Rm=100$, the LF
pulsation goes back to a single frequency. For this value of $Rm$, the
system is inside the instability pocket, where the vortices are
statistically stationary (red points in
Fig.\ref{fig:fig5}).

Such a Low Frequency pulsation has been reported both in experiments
and simulations with comparable values \cite{Araseki04}. In this last
reference however, the pulsation was associated with a loss of
homogeneity of the fluid in the azimuthal direction when the forcing
was non-axisymetric. The above results show that such a behavior may
be generated even in axisymetric configuration.

\section{Discussion and conclusion}

In this article, we have extended the numerical study presented in the
first part, describing an MHD flow driven by a travelling magnetic field
in an annular channel, and aiming to describe linear annular induction
electromagnetic pumps.  Here, we report results using larger values of
the flow Reynolds numbers, and also introducing the presence of
inlet/outlet boundary conditions in the axial direction, giving a
better description of experimental configurations.\\

For the turbulent case studied here, we have shown that the stalling
instability previously described for laminar flows takes the form of a
loss of homogeneity in the flow, for which a strongly negative
velocity appears in a given region of the channel. We have identified
this behavior as a local stalling of the flow, characterized by the
coexistence of two regions of fluids with very different velocity and
induced magnetic field. 

This is consistent with a simple model of several pumps operating in
parallel hydraulically, although not electrically in series. This
arises from the fact that, since the magnetic field is imposed only at
the boundaries, the simulations involve a local Hartmann number which
depends on the radial direction. As a consequence, the inner part of
the fluid always becomes unstable before the region of fluid close to
the outer cylinder, where the electrical currents are imposed. In this
case, the flow exhibits a Poiseuille-like profile close to the inner
cylinder, and an Hartmann-like profile close to the outer cylinder. 

The use of 2D simulations allowed us to explore a large region of the
parameter space $(Ha,Rm,Re)$. At large $Re$, the single marginal
stability curve observed for laminar flows is replaced by a pocket of
instability involving inhomogeneous velocity profiles. Although more
simulations are necessary to give a definitive conclusion, our results
suggest that the upper boundary of the instability pocket is
consistent with the scaling involving the Alfvenic Mach number  $M_b>$constant . As explained in
the first part of our article, this scaling law indicates that
magnetic flux expulsion is responsible for the loss of synchronism
close to the inner side of the pump, where the magnetic field, and
therefore the force, are weaker. 
It is important to note that an adverse pressure gradient is necessary for a pump to lie on the negative part of its characteristic curve. In our turbulent simulations, such a load is created by the inlet/outlet conditions at the ends of the pump.

Finally, consistently with previous experimental and numerical results
on annular linear induction EMPs, we have also observed the appearance
of a periodic modulation of the fields with a typical frequency
approximately $30$ times smaller than the applied external
frequency. This modulation takes here the form of axisymmetric
pulsating vortices which systematically occur close to the upper
marginal stability line of the reversed flow instability. \\

It is now interesting to compare these numerical results to existing
experimental configurations.  Medium size EMPs can reach magnetic
Reynolds numbers higher than $10$ and Hartmann numbers larger than
$800$ (with our definition of $Rm$ and $Ha$), while the largest EMPs
reported in the literature correspond to $Rm > 50$ and $Ha > 10^{5}$
\cite{Ota04,Araseki04}. This suggests that the axisymmetric loss of
homogeneity reported here may be relevant to experimental
configurations using similar types of forcing, in which the electrical
currents are applied only on one side of the pump. In this
perspective, it would be interesting to understand how this
intrinsically axisymmetric instability is modified by the presence of
electrical currents on both sides of the annular channel. Similarly,
secondary bifurcations towards non-axisymmetric states should be
expected in both single-sided or double-sided configurations, and the
study of this fully 3d states will be reported in a future work.\\

\begin{acknowledgments}
This work was supported by funding from the French program "Retour Postdoc" managed by Agence Nationale de la Recherche (Grant ANR-398031/1B1INP), and the DTN/STPA/LCIT of Cea Cadarache. 
The present work benefited from the computational support of the HPC resources of GENCI-TGCC-CURIE (Project  No.   t20162a7164)  and  MesoPSL  financed  by  the  Region
Ile  de  France  where  the present numerical simulations have been performed
\end{acknowledgments}

\bibliographystyle{jfm}
\bibliography{pemsim2}

\begin{thebibliography}
\bibitem

\bibitem{Galaitis76}
Gailitis A. and O. Lielausis, {\it Instability of homogeneous velocity distribution in an induction-type mhd machine}, Magnetohydrodynamics, {\bf 1},
69-79 (1976). Translation from Magnitnaya Gidrodinamika 1, 87-101 (1975).

\bibitem{Kirillov80}
Kirillov, I.R., Ogorodnikov, A.P., Ostapenko, V.P. ,{\it Local  characteristics  of  a
cylindrical  induction  pump  for  Rms larger than 1 } ,  English translation from Magnitnaya Gidrodinamika {\bf 2}, 107-€"113 (1980)

\bibitem{Karasev89} 
Karasev, B. G., Kirillov, I. R., Ogorodnikov, A. P.,
{\it 3500 m3/h MHD pump for fast breeder reactor} , Liquid Metal Magnetohydrodynamics.  Kluwer, Dordrecht, pp. 333-338
(1989).

\bibitem{Araseki00} 
Araseki, H., Kirillov, I., Preslitsky, G.,
  Ogorodnikov, A. P.,   {\it Double-supply-frequency  pressure  pulsation  in  annular
linear  induction  pump.  Part  I.  Measurement  and  numerical
analysis.} , Nucl. Engin. and Design, {\bf 195}, 85-100 (2000)
  
  \bibitem{Araseki04} 
Araseki, H., Kirillov, I., Preslitsky, G. V.,
  Ogorodnikov, A. P., {\it Magnetohydrodynamic instability in annular linear induction pump:: Part I. Experiment and numerical analysis} , Nucl. Engin. and Design, {\bf 227}, 29-50 (2004)
  
 \bibitem{Gonzales06}
Gonzales, M., Audit, E., Huynh, P.,  {\it HERACLES: a three-dimensional radiation hydrodynamics code} ,  Astronomy and Astrophysics, {\bf 464}, 429-435 (2006).

\bibitem{part1}
Gissinger, C., Rodriguez Imazio, P., Fauve, S., {\it Instability of electromagnetically driven flows, part I} Phys. Fluids (2015)

\bibitem{Ota04}
Ota, H., {\it Development of 160 m3/min large capacity sodium-immersed self-cooled electromagnetic pump}  , J. Nucl. Sci. Tech., {\bf 41}, 511-523 (2004)

\bibitem{Andreev78}
Andreev, A. M. \textit{et al},  {\it Results of an experimental investigation of electromagnetic pumps for the BOR-60 facility} ,  Magnitnaya Gidrodinamika, {\bf 1}, 617 (English translation) (1988)

\bibitem{Liu06}
Liu, W., Goodman, J., Ji, H.,  {\it Simulations of magnetorotational instability in a magnetized Couette flow} ,  Astrophys. J., {\bf 643}, 306 (2006).

\bibitem{Liu08}
W. Liu, {\it Numerical study of the magnetorotational instability in Princeton MRI experiment},  Astrophys. J., {\bf 684}, 515 (2008).

\bibitem{Gissinger11} 
Gissinger, C., Ji, H., Goodman, J. , {\it  The role of boundaries in the Magnetorotational instability} ,  Phys. Fluids.,  {\bf 24}, 074109 (2012).

\bibitem{Moffatt82}
Kamkar, H. \& Moffatt, H. K.,   {\it A dynamic runaway effect associated with flux expulsion in magnetohydrodynamic channel flow} ,  J. Fluid Mech.,  {\bf 121},107-122 (1982)
\end{thebibliography}


\end{document}